\documentclass[twocolumn,showpacs,preprintnumbers,amsmath,amssymb,prl]{revtex4}

\usepackage{graphicx}
\usepackage{dcolumn}
\usepackage{bm}

\begin{document}


\title{Thermal Radiation From Carbon Nanotube in Terahertz Range}

\author{A. M. Nemilentsau} \email{Andrei.Nemilentsau@gmail.com} \author{ G. Ya. Slepyan} \author{S. A. Maksimenko} 
\affiliation{Institute for Nuclear Problems, Belarus State University, Bobruiskaya 11, 220030, Minsk, Belarus}


\begin{abstract}
The thermal radiation from an isolated finite--length carbon nanotube (CNT) is theoretically investigated both in near-- and far-field zones.  The formation of the discrete spectrum in metallic CNTs in the terahertz range is demonstrated due to the reflection of strongly slowed--down surface--plasmon modes from CNT ends. The effect does not appear in semiconductor CNTs. The concept of CNT as a thermal nanoantenna is proposed. 
\end{abstract}


\maketitle

Carbon nanotubes --- quasi-one-dimensional carbon molecules --- have found
numerous real and potential applications as building blocks of nanoelectronic circuits \cite{circuit} and nanoscale optical elements \cite{Novotny}. Among others, the idea of CNT-based optical devices enabling control and enhancement
of radiation efficiency on the nanoscale, i.e. nanoscale antennas for infrared and visible light, is actively discussed \cite{Dresselhaus,Hanson,Slepyan_06,Burke}.  Noise properties and operational limits of such devices are substantially determined by the thermal fluctuations of electromagnetic field.  
Fundamental interest to the thermal radiation is also dictated by the ability of nanostructures to change the \textit{photonic local density of states} (LDOS), i.e. the electromagnetic vacuum energy \cite{Novotny, Gaponenko, Scully}. The effect  has been observed in microcavities, photonic crystals, nanoparticles in the vicinity of surface--plasmon resonances, etc. \cite{Novotny}. Thus, as the electromagnetic fluctuations are defined by the photonic LDOS, investigation of the thermal radiation is expected to bring new opportunities for the reconstruction of photonic LDOS in the presence of nanostructures.   In turn, the photonic LDOS is a key physical factor defining a set of well--known quantum--electrodynamical effects: the Purcell effect \cite{Scully}, the Casimir--Lifshitz forces \cite{Landau_Stat}, the electromagnetic friction \cite{Novotny}, etc. All aforesaid  stimulates investigation of thermal radiation on nanoscale. 

During last decade,  there has been considerable interest to optical properties of surface--plasmon structures  \cite{Pitarke, Girard}.  In particular, spherical gold particles in the vicinity of the plasmon resonance have been shown to be an effective nanoantennas \cite{Kuhn}. Recently, the strong coherent coupling between individual optical emitters and
guided plasmon excitations has been predicted \cite{Chang} introducing thus quantum optics of nanoplasmonic structures. 
Thermal radiation  in systems with surface plasmons is influenced by the near--field effects and is known to be  considerably different from the black--body radiation \cite{Carminati, Shchegrov}. Earlier theoretical studies of CNTs showed the existence of  low--frequency plasmon branches \cite{Lin, Longe} and the formation in CNTs of strongly slowed-down electromagnetic surface waves \cite{Slepyan_99}. Such waves define pronounced Purcell effect in CNTs \cite{Bondarev} and potentiality of CNTs as Cherenkov-type emitters \cite{Batrakov}. Geometrical resonances --- standing surface waves excited due to the strong reflection from the tips of finite--length CNTs --- qualitatively distinguish CNTs from the planar structures investigated in \cite{Carminati, Shchegrov}. One can expect an essential role of these resonances in the formation of the CNT's thermal radiation. The black--body radiation distribution from multi-walled CNT bundles  reported in Ref. \cite{Li} has been observed above the frequency range of surface plasmon modes and, what is more, is influenced by the nonhomogeneous broadening due to the CNT length and radius dispersion and multi-walled effects.   

In this Letter we report calculations of the CNT thermal radiation spectra in the terahertz frequency range and demonstrate  spatial coherence of the radiation. The model of the CNT optical properties based on the effective boundary conditions \cite{Slepyan_99} and the fluctuation--dissipative theorem in the Callen--Welton form \cite{Landau_Stat} have been utilized in our analysis. We show that the CNT thermal radiation is substantially different from the black--body radiation \cite{Landau_Stat} and the radiation of planar structures with surface plasmons \cite{Carminati, Shchegrov}. 

Consider an isolated single--wall CNT of the cross--sectional radius $R_{CN}$ and the length $L$, aligned along the $z$ axis of a cylindrical coordinate basis $(\rho,\phi,z)$ with the origin in the CNT geometrical center. We restrict consideration to the case  $R_{CN}<<2\pi/k$, which implies the incident field to be slowly varied within the CNT cross-section; here $k=\omega/c$, $\omega$ is the electromagnetic field frequency, and $c$ is the vacuum speed of light. The CNT is assumed to be placed into an optically transparent rarefied medium; the CNT is in the thermal equilibrium with this medium. Optically, this medium is equivalent to vacuum. The  Hamiltonian gauge, which implies the scalar potential to be equal to zero, is used for electromagnetic field. Then, to calculate thermal fluctuations of the field in the presence of CNT we make use the fluctuation--dissipative theorem, which expresses the correlations of the vector potential in terms of the retarded Green tensor  $\underline{\mathbf{G}}(\mathbf{r}_1,\mathbf{r}_2,\omega)$ of a system \cite{Landau_Stat}: 
\begin{equation}
\left<A_{n}(\mathbf{r}_1)A_{m}^*(\mathbf{r}_2)\right>_{\omega} = \left[\hbar+\frac{2\Theta(\omega,T)}{\omega}\right]  \textrm{Im}\left[G_{n m}(\mathbf{r}_1,\mathbf{r}_2,\omega)\right]\,, \label{Correlator_potential}
\end{equation}
where $n,m=\rho,\phi,z$, $\Theta(\omega, T) = \hbar\omega/\left[\textrm{exp}(\hbar\omega/k_B T)-1\right]$, $T$ is the temperature, $\hbar$ and $k_B$ are the Planck and Boltzmann constants, respectively. 

The equation $\underline{\mathbf{G}}^{(0)}(\mathbf{r}_1,\mathbf{r}_2,\omega) = (\underline{\mathbf{I}} + k^{-2}\nabla_{\mathbf{r}_1} \otimes \nabla_{\mathbf{r}_1} ) \, G^{(0)}(\mathbf{r}_1,\mathbf{r}_2,\omega)$ with  $\underline{\mathbf{I}}$ as the unit tensor and $\nabla_{\mathbf{r}_1} \otimes \nabla_{\mathbf{r}_1}$ as the operator dyadic acting on variables $\mathbf{r}_1$, relates the retarded free--space Green tensor with the corresponding scalar Green function $G^{(0)}(\mathbf{r}_1,\mathbf{r}_2,\omega)=\exp{(i k |\mathbf{r}_1-\mathbf{r_2}|) }/|\mathbf{r}_1-\mathbf{r_2}|$. 
For a given $m$, the column  $G_{n m}^{(0)}(\mathbf{r}_1,\mathbf{r}_2,\omega)$ can  formally  be considered  as a field vector induced in the point $\mathbf{r}_1$ by a source located in the point $\mathbf{r}_2$. The scattering problem for that incident field is formulated by the equation $\bigl(\nabla_{\mathbf{r}_1} \times \nabla_{\mathbf{r}_1}\times - k^2\bigr)\,\underline{\mathbf{G}}(\mathbf{r}_1,\mathbf{r}_2,\omega)=4\pi \underline{\bf I} \,\delta(\mathbf{r}_1-\mathbf{r}_2)$ imposed to the effective boundary conditions on the CNT surface \cite{Slepyan_99}. We present the solution in the form of the simple--layer potential \cite{Colton_b1983}:  
\begin{eqnarray}
&&G_{n m}(\mathbf{r}_1,\mathbf{r}_2,\omega)=G^{(0)}_{n m}(\mathbf{r}_1,\mathbf{r}_2,\omega)+\frac{i \omega R_{CN}}{ c^2}  \cr 
&&\qquad\times \int\limits_{-L/2}^{L/2}j_z^{(m)} (z;\mathbf{r}_2) \int\limits_{0}^{2\pi} G_{n z}^{(0)}(\mathbf{r}_1,\mathbf{R},\omega)  d\phi   \, dz\,,  \label{Dyson_equation}
\end{eqnarray}
where $j_z^{(m)} (z;\mathbf{r}_2)$ is the normalized  density of the axial current induced on the CNT surface by the incident electric field $G_{z m}^{(0)}(\mathbf{R},\mathbf{r}_2,\omega)$, $\mathbf{R}=\{R_{CN},\phi,z\}$. While deriving (\ref{Dyson_equation}) we assumed the incident field source distant from the CNT  farther than its radius;  therefore we can neglect  the current $j_z^{(m)}$ dependence on the  azimuthal variable $\phi$. 

Eq. (\ref{Dyson_equation}) with arbitrary $j_z^{(m)}$ satisfies the afore stated equation for the retarded Green tensor  and the radiation condition at $|\mathbf{r}_1-\mathbf{r}_2|\to\infty$. Using the effective boundary conditions  \cite{Slepyan_06},  we obtain for $j_z^{(m)} (z;\mathbf{r}_2)$ the integral equation as follows
\begin{eqnarray}
\int_{-L/2}^{L/2} j_z^{(m)}(z';\mathbf{r}_2) {\cal K}(z-z') dz' + C_1 e^{-i k z} + C_2 e^{i k z} \nonumber \\ 
= \frac{1}{2\pi}\int_{-L/2}^{L/2} {\cal K}_1 (z-z') \int_{0}^{2\pi} G_{z m}^{(0)}(\mathbf{R}',\mathbf{r}_2,\omega) d\phi ' dz', \label{integral_equation}
\end{eqnarray}
where
\begin{eqnarray}
{\cal K}(z)=\frac{1}{\sigma_{zz}(\omega)} {\cal K}_1 (z)- \frac{2 i R_{CN}}{\omega} \int_0^{\pi} \frac{e^{i k r}}{r} \, d\phi\,, \label{kernel}
\end{eqnarray}
$r=\sqrt{z^2+4 R_{CN}^2 \sin^2(\phi/2)}$, ${\cal K}_1 (z)= \textrm{exp}(i k |z|)/2 i k$, $\sigma_{zz}(\omega)$ is the CNT axial conductivity  \cite{Slepyan_99}, $C_1$ and $C_2$ are constants determined by the edge conditions $j_z^{(m)}(\pm L/2;\mathbf{r}_2)=0$ \cite{Slepyan_06}. The index $m$ and the variable $\mathbf{r}_2$ appear  in Eqs. (\ref{Dyson_equation}) and (\ref{integral_equation}) only as parameters. Note that these equations, as they couple the Green tensor of the system considered  and the free--space Green tensor, play the role of the Dyson equation for CNTs. 

It is of importance that the role of scattering by CNT \textit{is not reduced} to a small correction to the free--space Green tensor. This means that the iteration procedure conventionally  used for the Dyson equation solving gets inapplicable in our case. Because of that, the direct numerical integration of Eq. (\ref{integral_equation}) has been performed with integral operators approximated by quadrature formula and subsequent transition to a matrix equation. 
 
First term in square brackets in Eq. (\ref{Correlator_potential}) is due to zero point energy of electromagnetic field and is further omitted. The second term comprises the CNT's thermal radiation and the black--body radiation of the surrounding medium. The latter one  is represented by the correlator $ \bigl<A_{n}^{(B)}(\mathbf{r}_1)A_{m}^{(B)*}(\mathbf{r}_2)\bigr>_{\omega}\equiv D^{(B)}_{n m}(\mathbf{r}_1,\mathbf{r}_2,\omega)$  of the black--body radiation vector potential $\mathbf{A}^{(B)}$. To separate off the CNT's intrinsic thermal radiation,  we make use the method presented in Ref. \cite{Landau_Stat} (see the problems after Sect. 77). The electric field intensity of the CNT's thermal radiation $I_{\omega}(\mathbf{r}_0)=|\mathbf{E}(\mathbf{r}_0)|^2$ is given by 
\begin{equation}
I_{\omega}(\mathbf{r}_0) = k^2 \sum_{n=1}^3\bigl[\bigl<|A_n(\mathbf{r}_0)|^2\bigr>_{\omega}  -D_{nn}^{(B)}(\mathbf{r}_0,\mathbf{r}_0,\omega)\bigr]\,.  \label{electric_intensity}
\end{equation}
Here, the relation $E_n=-i k A_n$ has been accounted for.  Note that Eq. (\ref{electric_intensity}) also determines the intensity of the electric field fluctuations in the case when the CNT temperature is much higher than the temperature of the surrounding medium. 

By analogy with Eq. (\ref{Dyson_equation}), the vector potential $A_n^{(B)}(\mathbf{r}_1)$ is written as 
\begin{eqnarray} \label{Vector_potential}
&&A_n^{(B)}(\mathbf{r}_1)=A_n^{(0)}(\mathbf{r}_1) \nonumber \\
&&\qquad+\frac{ R_{CN} }{ c }\int \limits_{-L/2}^{L/2} j(z) \int\limits_{0}^{2\pi}G_{n z }^{(0)}(\mathbf{r}_1, \mathbf{R},\omega)\, d\phi \,dz\,,
\end{eqnarray}
where $A_n^{(0)}(\mathbf{r})$ is the vector potential of the  free--space black--body radiation (i.e., of the black--body radiation in the absence of CNT). Second term in (\ref{Vector_potential}) describes scattering of the free--space black--body radiation by CNT. The density of the current $j(z)$  induced on the CNT surface is found as a solution of Eq. (\ref{integral_equation}) with $i\omega A^{(0)}_z(\mathbf{R},\omega)/c$ substituted instead of $G_{z m}^{(0)}(\mathbf{R},\mathbf{r}_2,\omega)$. 
To calculate $D_{n m}^{(B)}$ we utilize (\ref{Vector_potential}) and take into account that the correlator $\bigl<A_{n}^{(0)}(\mathbf{r}_1)A_{m}^{(0)*}(\mathbf{r}_2)\bigr>_{\omega}$ is defined by Eq.  (\ref{Correlator_potential}) with the free--space Green tensor $G_{n m}^{(0)}(\mathbf{r}_1,\mathbf{r}_2,\omega)$ in the right--hand part. 

\begin{figure}[!htb]
\begin{center}
\includegraphics*[width=3.0 in]{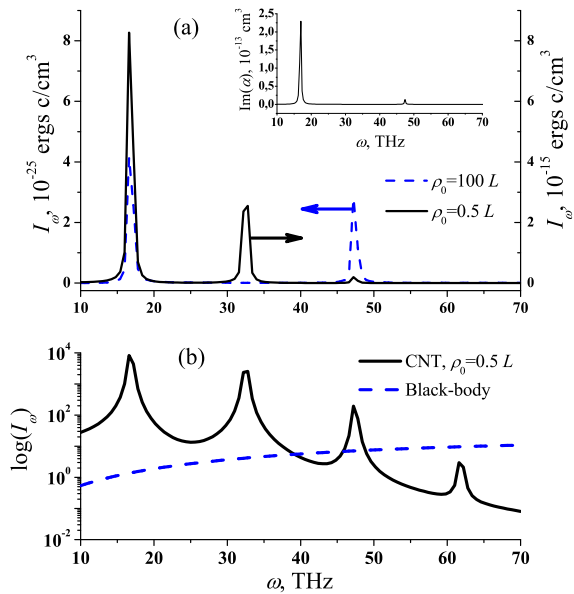}
\end{center}
\caption{$(a)$ Thermal radiation spectra of metallic (15,0) CNT in the cross-section  $z_0=0$ and at the distance from the CNT axis $\rho_0=100 L$,  (dashed line, left ordinate axis) and $\rho_0=0.5 L$ (solid line, right ordinate axis). On the inset, the polarizability of the CNT  is depicted.  $(b)$ Thermal radiation from CNT (solid line) and black--body radiation (dashed line) in the near--field zone. In both pictures $L=1\,\mu$m and $T=300$ K. The relaxation time  $\tau = 3\times 10^{-12}$ s has been used under calculation of the axial conductivity $\sigma_{zz}$. }  \label{fig1}
\end{figure}

The thermal radiation spectra  from (15,0) CNT at different distances from the CNT axis are depicted in Fig. \ref{fig1}a; Fig. \ref{fig1}b presents one of the spectra in the logarithmic scale. 
For comparison, the black--body radiation intensity $I^{(B)}_{\omega}(\mathbf{r}_0)=4\omega^2 \Theta(\omega,T)/c^3$ is also depicted. 

The spectrum depicted in Fig. \ref{fig1}a demonstrates a number of equidistant discrete spectral lines with decreasing intensities superimposed by the continuous background.  Such a structure  is inherent to spectra both in the far--field (dashed line) and near--field (solid  line) zones. The peculiarity of the near--field zone is the presence of additional spectral lines absent in the far--field zone. Thus, the thermal radiation spectra presented in the figure are qualitatively differ from both  black-body radiation \cite{Landau_Stat} and radiation of semi--infinite SiC samples \cite{Shchegrov}. In the latter case, the discrete spectrum is observed only  in the near--field zone \cite{Shchegrov}.  

Comparison of the thermal radiation and the CNT's polarizability spectra depicted in Fig. \ref{fig1}a 
reveals the coincidence  in the far--field zone of the thermal radiation resonances and the polarizability resonances. The latest are the \textit{dipole} geometrical resonances of surface plasmons \cite{Slepyan_06} defined by the condition $\textrm{Re}[\kappa(\omega)] L \cong \pi (2s - 1)$ with $\kappa(\omega)$ as the plasmon wavenumber; $s$ is a positive integer. It should be noted that the polarizability (and the thermal radiation) resonances is found to be significantly shifted to the red as compared to the perfectly conducting wire of the same length because of the strong slowing-down of surface plasmons in CNTs:  $\textrm{Re}[\kappa(\omega)]/k \approx 100 $ \cite{Slepyan_99}. In particular, for $L=1\,\mu$m, the geometrical resonances fall into the terahertz frequency range. The attenuation is small in a wide frequency range below the interband transitions. Additional spectral lines in the near--field zone are described by the condition $\textrm{Re}[\kappa(\omega)] L \cong 2\pi s$ and are due to the \textit{quadrupole} geometric resonances.  Thus, the resonant structure of the thermal radiation spectra is entirely determined by the finite--length effects. 
Note that similar structure of the thermal radiation spectra is predicted for 2D--electron gas \cite{Richter}. Resonances in Ref. \cite{Richter} are due to excitations of other physical nature ---optical-phonon modes of the barrier material.   

The presence of singled out resonances 
illustrated by Fig. \ref{fig1}a allows us to propose metallic CNTs as far--field and near--field \textit{thermal antennas of a new type} (thermal antennas based on photonic crystals have recently been  considered in Ref. \cite{Laroche}). Taking into account the high temperature stability of CNTs, the CNT thermal antennas can be excited by Joule heating from the direct (low--frequency) electric current.  

Accordingly to Refs. \cite{Hanson,Slepyan_06,Burke} the maximal efficiency of vibrator CNT--antennas is reached at frequencies of the surface--plasmon dipole resonances. Figure \ref{fig1}a  shows that the intensities of spectral lines of the thermal radiation go down with the resonance number much slowly  than the polarizability peaks. This means that the signal/noise ratio for the CNT-based antennas is maximal for the first resonance and decreases fast with the resonance number. 
\begin{figure}[!htb]
\begin{center}
\includegraphics*[width=3.0 in]{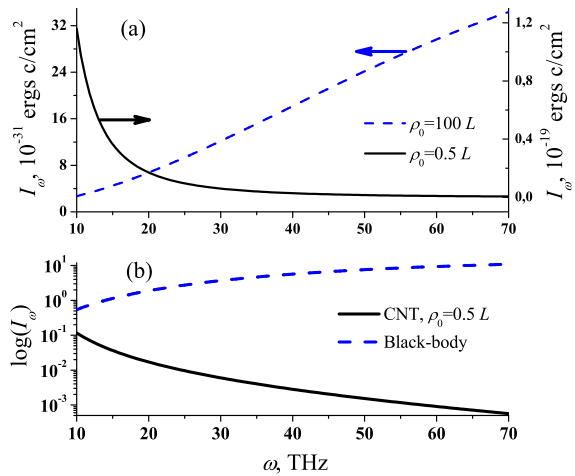}
\end{center}
\caption{Same as Fig. \ref{fig1}, but for the (23,0) semiconducting zigzag CNT.}  \label{fig2}
\end{figure}

As different from metallic CNTs, semiconducting ones do not reveal isolated resonances  in both  far--field and near--field zones, see Fig. \ref{fig2} as an illustration.  Such a peculiarity can easily be understood by accounting for the strong attenuation of surface plasmons  in semiconducting CNTs  whereas the slowing-down remains of the same order. That is why in this case the Q--factor of geometrical resonances  turns out to be substantially smaller and they do not manifest themselves as separated spectral lines. In the same way, the thermal radiation intensity of semiconducting CNTs is substantially smaller than that of metallic ones and displays qualitatively different spectral properties  in the near--field zone: monotonous growth of the intensity with frequency inherent to metallic CNTs  changes into  monotonous declining (see Fig. \ref{fig2}a). As the thermal  spectra are strongly dependent on the CNT conductivity type and length, the near-field thermal radiation spectroscopy proposed in \cite{Shchegrov} for testing the surface--plasmon structures, can be expanded to CNTs.

Figure \ref{fig2}b demonstrates that in the frequency range considered  the black-body radiation intensity considerably exceeds the thermal radiation of semiconducting CNTs: $I_{\omega} \ll I_{\omega}^{(B)}$. In the vicinity of geometrical resonances the same property is inherent  to metallic CNTs, see Fig. \ref{fig1}b. This means that CNTs as  building blocks for nanoelectronics and nanosensorics possess uniquely low thermal noise and, thus, provide \emph{high electromagnetic compatibility on the nanoscale}: 
Their contribution to the electromagnetic fluctuations in nanocircuits is negligibly small as compared to the contribution of dielectric substrate. 
More generally, the later example illustrates peculiarity of the electromagnetic compatibility problem on the nanoscale  motivating future research investments into the problem.  

\begin{figure}[th]
\begin{center}
\includegraphics*[width=3.0 in]{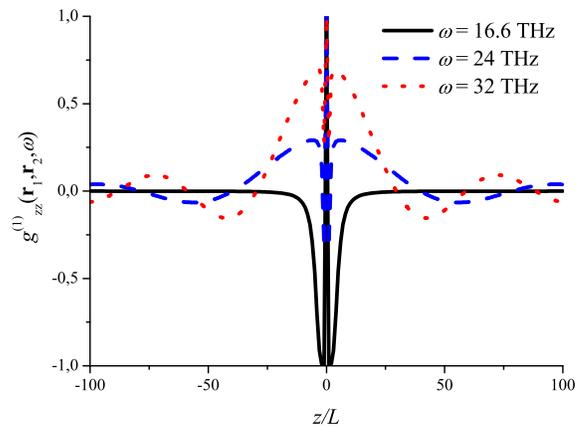}
\end{center}
\caption{First-order correlation function $g_{zz}^{(1)}(\mathbf{r}_1,\mathbf{r}_2,\omega)$ for metallic (15,0) CNT. $\mathbf{r}_1=\{0.5L, \phi, z\}$, $\mathbf{r}_2=\{0.5L, \phi, 0\}$, and $\phi$ has an arbitrary value. }  \label{fig4}
\end{figure}

In vie of the CNTs' application as tips for near-field optical microscopy \cite{Novotny}, we have studied the  spatial structure of the electromagnetic fluctuations near CNT,  characterized by the normalized first--order correlation tensor
\begin{equation}
g_{nm}^{(1)}(\mathbf{r}_1,\mathbf{r}_2,\omega)= \frac{\langle A_n(\mathbf{r}_1) A_m^*(\mathbf{r}_2)\rangle _{\omega}}
	{\sqrt{\langle |A_n(\mathbf{r}_1)|^2\rangle _{\omega}\langle |A_m(\mathbf{r}_2)|^2\rangle _{\omega}}}. \label{correlation}
\end{equation}
Calculations of the axial--axial component of this tensor  in the near-field zone are presented in Fig. \ref{fig4}. Because of the strong slowing--down of surface plasmons in CNTs, the vacuum wavelength  of the dipole geometrical resonance  $\lambda \sim 100 L$ exceeds significantly the corresponding coherence length, which is estimated from  the peak width in Fig. \ref{fig4} (solid line) to be  about $10L$. At the quadrupole resonance frequency (dotted line) and far away from resonances (dashed line) the coherence length considerably increases. Physical interpretation of such a behavior is analogous to that given in  Ref. \cite{Carminati} for planar structures: The coherence length is of the order of magnitude of the spatial scale of the dominant mode. At the dipole resonance the dominant mode is surface plasmon those spatial scale is dictated by  $\kappa^{-1} \ll 2\pi \lambda$. Far away from the resonance the contribution of volume modes considerably increases extending the spatial scale to $\approx \lambda/2$. 																	

Concluding, the thermal radiation from single--wall finite-length CNTs in the terahertz range has been theoretically investigated. Discrete spectral lines in the radiation spectra of metallic CNTs, originated from the geometrical antenna resonances, is predicted to exist in both near-- and far--field zones. The effect allows a concept of metallic CNT as a thermal antenna and is of importance for the  CNT--spectroscopy, the nanoantenna design, the high--resolution near--field optical microscopy and  the thermal noise control in nanocircuits.

The research was partially supported by the INTAS (grants
03-50-4409 and 05-1000008-7801), and the Belarus Foundation for Fundamental Research and Russian
Foundation for Basic Research (grant F06R-101). AMN acknowledges PhD fellowship INTAS grant 05-109-4595.

\end{document}